\documentstyle[epsf]{l-aa}     
\begin{document}

\title{The accuracy of parameters determined with the core-sampling method:
application to Voronoi tessellations}
   \subtitle{}

   \thesaurus{02         
              (12.03.4   
               12.12.1   
               03.13.6)   
}
   \author{Andrei G. Doroshkevich \inst{1}, Stefan Gottl\"ober
   \inst{2}, and S{\o}ren Madsen\inst{3}}
   \institute{
       Theoretical Astrophysics Center,
       Juliane Maries Vej 30,
       DK-2100 Copenhagen \O, Denmark
  \and
       Astrophysikalisches Institut Potsdam,
       An der Sternwarte 16,
       D-14482 Potsdam,
       Germany
   \and
       Copenhagen University, Astronomical Observatory,
       Juliane Maries Vej 30,
       DK-2100 Copenhagen \O, Denmark
}

   \date{Received -- -- --; accepted -- -- --}

   \maketitle
   \markboth{Doroshkevich et al.}{Core-sampling}

\newcommand{\apj}{ApJ}
\newcommand{\mnras}{MNRAS}
\newcommand{\Mpc}{\hbox{Mpc}}

\begin{abstract}

The large-scale matter distribution represents a complex network of
structure elements such as voids, clusters, filaments, and sheets. This
network is spanned by a point distribution.  The global properties of
the point process can be measured by different statistical methods,
which, however, do not describe directly the structure elements.

The morphology of structure elements is an important property of the
point distribution. Here we apply the core-sampling method to various
Voronoi tessellations. Using the core-sampling method we identify one-
and two-dimensional structure elements (filaments and sheets) in these
Voronoi tessellations and reconstruct their mean separation along
random straight lines.  We compare the results of the core-sampling
method with the a priori known structure elements of the Voronoi
tessellations under consideration and find good agreement between the
expected and found structure parameters, even in the presence of
substantial noise. We conclude that the core-sampling method is a
potentially powerful tool to investigate the distribution of such
structure elements like filaments and walls of galaxies.

\end{abstract}

\section{Introduction}

The first deep galaxy surveys which have become available have shown that
the galaxies are not at all homogeneously distributed on large scales.
Large voids, small filaments,  massive clusters, and walls of galaxies can
be observed. How these structures have evolved from the initially nearly
homogeneous universe is one of the central problems in modern
cosmology. Many theoretical models of structure formation have been
suggested. Powerful mathematical statistical methods are necessary in
order to describe this structure, to obtain its quantitative
characteristics, to compare the theoretical models and simulations
with observational catalogues, and, eventually, to find the correct
theory of structure formation.

The standard approach for testing models is to define a point process
which can be characterized by statistical methods. This could be the
distribution of galaxies of a specific type in deep surveys or
clusters of galaxies. In order to compare models of structure
formation,  the different distribution of dark matter particles in
N-body simulations could be tested as well.

The most widely used statistics are the n-point correlation functions,
the counts-in-cells method, and the void probability function.  The
geometrical and topological properties of the point distribution can
be investigated by the percolation technique (Zeldovich, Einasto,
Shandarin 1982), the minimal spanning tree (Barrow, Bhavsar, \& Sonoda
1985), the genus of the smoothed density field (Gott et al. 1986,
1989), and the Minkowski functionals (Mecke et al. 1994).

All these different statistics measure the global properties of a
point process. However, when looking at a point process we also see
different structures. The morphology of these structures is an
important property of the point distribution. (Note also the warning
of Barrow \& Bhavsar (1987) that we tend to see by eye structures
which do not exist). Therefore, it is important to develop statistics
which distinguish between different structure elements in a point
distribution and find their typical scales (if any exists).

In this direction a first attempt was made by Vishniac (1986). He used
the moments of a point distribution in a window to measure the amount
of filaments in two-dimensional galaxy distributions. Recently, this
method was generalized to three-dimensional point distributions (Luo,
Vishniac 1995). Both the number of filaments and walls and typical
scales of the structure can be found. Minkowski functionals are also
efficient discriminators for idealized one-, two- or three-dimensional
structure elements (Schmalzing, Kerscher, Buchert 1996). However,
using Minkowski functionals it is difficult to extract such structure
elements from a superposition of all possible elements including
Poisson noise.

Using the core-sampling method (Buryak, Demia\'nski, \& Doroshkevich
1991), Buryak, Doroshkevich \& Fong 1994) one can find structure
elements and their typical separation. The method was designed to find
one-dimensional filaments and two-dimensional walls within
observational surveys of galaxies (Buryak, Doroshkevich \& Fong, 1994,
Doroshkevich et al. 1996b) and in simulated samples (Doroshkevich et
al. 1996a). Here we want to apply the method to an idealized model
mimicking many features of the observed patterns in the galaxy
distribution. The model employed here is based on the concept of a
Voronoi tessellation (Goldwirth, da Costa, \& Van de Weygaert
(1995)). Such a test allows us to compare the input and output
structure parameters and, thus, to test how powerfully this method can
discriminate structure elements, determine their distribution and
typical parameters.

The basic idea of the core-sampling method is to reduce the analysis
of a three-dimensional galaxy distribution to the investigation of the
distribution of structure elements along random straight lines (the
'core'). Such an approach allows us to avoid any discussions and
descriptions of the very complicated multiconnected structure as a
whole as well as any conventional definition of a 'structure element'
and 'void' in the three-dimensional space. The core-sampling method
has to deal only with a one-dimensional point distribution. It allows
us to define two typical populations of structure elements, namely,
filaments and sheets, and provides us with two objective
characteristics of the spatial distribution of these elements, namely
the surface density of filaments and the linear density of sheets.
The surface density of filaments $\sigma_f$ is simply the mean number
of filaments intersecting a unit area of arbitrary orientation, while
the linear density of sheets $\sigma_s$ is simply the mean number
density of sheets crossing an arbitrary straight line.  These
characteristics provide a local description of a random
three-dimensional network structure. In Sect. 2 we will briefly
outline the general properties of this method

The three-dimensional Voronoi tessellation consists of a number of
Voronoi cells enclosed by the planes bisecting the lines between the
nuclei of neighbouring cells (van de Weygaert, 1991). The intersections
of these planes are lines and points. Therefore, the three-dimensional
Voronoi tessellations consists of three structure elements: the walls
around the voids, the edges (intersection of walls) and the
nodes. Within the three-dimensional Voronoi tessellation particles may
be distributed on each of these structure elements or distributed
around them according to a given distribution function. Moreover, the
Voronoi tessellation can be superimposed by randomly distributed
particles. Thus it is a unique tool for testing algorithms purporting
to describe the structure elements of a point distribution. 

In Sect. 3 we will briefly describe Voronoi tessellations and the
creation of synthetic samples. We apply the core-sampling method to
these synthetic samples and compare the results with the input
structure element. We discuss the accordance between the expected and
found structure parameters and conclude that the core-sampling method
is a powerful tool to investigate the distribution of structure
elements.

\section{The Core-Sampling Method}

\subsection{Basis of core-sampling method}

The mathematical basis for this method had been described by Buryak,
Demia\'nski, \& Doroshkevich (1991) and Buryak, Doroshkevich \& Fong
(1994). A recent detailed discussion can be found in Doroshkevich et
al. (1996b).

First let us  give the salient points of the core-sampling method.

\begin{enumerate}

\item The distribution of structure elements along a random straight
      line is assumed to be Poissonian. Thus, a 1D cluster analysis
      is utilized to discriminate the structure elements among the
      sample of points and to find the number and the mean separation
       of structure elements.

\item For the following cluster analysis the fields are then all 
      organized into an
      `equivalent single field' by combining the separate 1D
      distributions one after the other along a line, with the first
      point of a field placed on top of the last point of the
      preceding field.

\item The dependence of the number and mean separation of structure
      elements on the diameter of the core allows a rough
      discrimination between the filament and sheet-like populations of the
      structure elements and yields the fundamental characteristics
      of the structure, namely, the surface density of filaments,
      $\sigma_f$, and the linear density of sheets, $\sigma_s$.

\item The sample of points under consideration  is reduced by rejecting
      poorer and sparser structure elements.
      In this manner, the mean characteristics can be found
      as a function of the threshold richness.
\end{enumerate}

Let us emphasize here that in practice the one-dimensional analysis is
very convenient in many respects.  We are using here and in numerical
simulations cylinders around straight lines, but they can be easily
replaced by cones for observational surveys. Thus, the core-sampling
method can be directly used to analyze pencil beam surveys as well as
almost two-dimensional samples like the slices of the deep Las
Campanas survey or real three-dimensional surveys. Moreover, in case
of two- and three-dimensional samples it allows measurement of
structure parameters in different directions, for instance, both along
the line of sight and the transverse direction. Thus, redshift space
distortions can be extracted (Doroshkevich et al. 1996b).

The assumption of a functional form for the distribution of structure
elements along the core is essential for our method, because it
enables transformation of the point distribution into a distribution of
structure elements.  Once this transformation is accomplished, the
appropriateness of the assumed functional form can be tested and the
values of the functional parameters can be determined.

As stated above, we assume a 1D Poissonian law for the distribution of
structure elements (not galaxies) along the axis of cylinder.  It is a
simple distribution, and we shall see that it is also a reasonable
assumption.  Indeed, the Poissonian distribution arises naturally for
some theoretical models (White 1979, Buryak et al. 1991) when the mean
separation of structure element exceeds the correlation length.  The
validity of this assumption, however, cannot be tested a priori, and a
possible difference between the assumed and the actual distributions
depends on the sample in question, thus limiting the precision of the
final results. This assumption was valid for all cases in previous
investigations. Here we are testing the method with Voronoi
tessellations, for which our assumption has to be tested as well.

\subsection{Sample preparation}

The first step of core-sampling analysis is the preparation of a
sample with suitable parameters. This sample depends both on the
radius of the cylinder $R_c$ and the linking length $l_{c}$ of the
culling procedure.  The sample is the basis for the further analysis.

The diameter of the cylinder must be a few times smaller than the
expected size of the cells in order to avoid a masking effect due to
possible overlapping of projected elements. On the other hand, the
cylinder must be wide enough to allow during further analysis the
sequential reduction of the radius by a factor of at least 2,
preferably more, because the radius is that diagnostic parameter which
allows one to distinguish between filaments and sheets.  The number of
particles has to be large enough to guarantee stability and
reliability of results even for the smallest radius used for analysis.
In practice one observes that this can be achieved if a structure
element contains, in the mean, at least 5, or preferably more,
particles.

The final step of the sample preparation is sample reduction, which
is performed sequentially during the analysis. Sample reduction
provides an additional method of distinguishing between different
populations of filaments and sheets according to the mean density or,
in other words, according to the multiplicity of the structure
elements. It has to be performed before any further steps of analysis.

To do the sample reduction, a 1D cluster analysis of each field is
performed with decreasing linking length $l_{c}$ of the culling
procedure. All clusters with fewer members than a constant threshold
multiplicity $\mu_{thr}$ are rejected. Then the remaining number of
particles in the sample $N_r$ depends on the current linking length
$l_c$ and the threshold $\mu_{thr}$. For each linking length $l_c$ a
certain fraction $f = N_r/N_a$ of points remains ($N_a$ is the total
number of particles). It is convenient to use the fraction $f$
together with the threshold multiplicity $\mu_{thr}$, as parameters
characterizing the reduced sample.

This procedure reduces the full sample, rejecting more and more points
associated with poorer or sparse clumps.  Only the tighter clumps are
retained for further analysis. This approach emphasizes the local
density within the structure elements. It allows the rejection of the
low-density haloes from the structure elements and sparse structure
elements as a whole.

\subsection{Detection of the mean separation of structure elements}

The next step of analysis is the detection of the mean separation of
structure elements.  At first, the axes of all cylinders are randomly
combined to one line by identifying the last point of one axis with
the first of the next.  At this line a 1D cluster analysis with
increasing linking length $R$ is performed. Assuming a Poisson
distribution, the number of clusters $N_R$ as a function of $R$ is
given by the relation
\begin{equation}
\ln(N_R) = \ln(N_0) - R/R_0.
\label{lnN}
\end{equation}
For a true Poissonian sample, these values are related to the length of
the sample, $D_0$, defined by the nearest and the farthest points in
an average field, via the relationship
\begin{equation}
N_0/D_0 = 1/R_0.
\label{ND}
\end{equation}
Thus, the difference between the linear density of structure elements
$N_0/D_0$ and their mean separation $R_0$ obtained from
Eqs. (\ref{lnN}) and (\ref{ND}) can be considered a measure of the
systematic error due to the deviation of the actual from the assumed
(Poissonian) distribution of structure elements along the analyzed
cylinder. To decrease this error we use an automatic procedure to find the
optimal interval for $R$ for the fit to Eq. (\ref{lnN}).  The upper
limit to this interval is fixed by the input of a minimum number
$N_R\geq N_{min}$, whereas the lower limit can vary and is defined by
the condition
\begin{equation}
1-\epsilon \leq R_0 N_0/D_0 \leq 1+\epsilon
\end{equation}
where $\epsilon$ is a desirable precision of fitting parameters.
In general, a reasonable precision ($\approx$10\%) can be achieved
in the analysis.

If a desirable precision cannot be achieved, the reason is usually
that the distribution used is far from Poissonian and the sample under
consideration must be changed (e.g., by changing the cylinder radius).

\subsection{Identification of structure elements}
The final step of analysis is the discrimination of filaments and
sheet-like structure elements and determination of the parameters
$\sigma_f$ and $\sigma_s$ for both populations. To this end, we use a
simple geometrical model for the structure elements (Buryak,
Doroshkevich \& Fong (1994)). According to this model, in each narrow
cylinder the structure can be considered as a system of randomly
distributed lines (filaments) and planes (sheets) which contain all
points.  In this model, the filaments are considered straight lines,
and the sheets are considered flat planes.  Of course, this approach
is limited, and it cannot be used as an accurate description of the
true matter distribution on large scales; one can best characterize it
as an intermediate step between the local description of the matter
distribution with density and velocity fields and the global
description obtained by the topology or Minkowski techniques.

As it was mentioned above, we characterize the random distribution of
straight lines (filaments) by their surface density, $\sigma_f$, i.e.,
the mean number of lines intersecting an unit area of arbitrary
orientation.  The random distribution of planes can be characterized
by the linear density, $\sigma_s$, i.e., the mean number density of
planes (sheets) crossing an arbitrary straight line. Equivalently, we
can use the values $D_s = \sigma_s^{-1}$ and $D_f = \sigma_f^{-1/2}$
as typical measures of the mean separation of structure elements. To
characterize the structure as a whole we can estimate the mean
separation of structure elements --- filaments and sheets combined ---
by $<D_{fs}>$, which can be thought of as the diameter of a sphere
containing, on average, two structure elements,
\begin{equation}
<D_{fs}> = {2\over{\sigma_s+\sqrt{\sigma_s^2+\pi\sigma_f}}}
\end{equation}

The main characteristics of the structure can be obtained by
fits to the radial dependence of the linear density of clusters,
$N_0/D_0$, and to the radial dependence of the mean separation of
structure elements, $R_0$,
\begin{equation}
N_0/D_0 = \sigma_s + \pi~R_c\sigma_f,
\label{rd1}
\end{equation}
and
\begin{equation}
1/R_0 = \sigma_s + \pi~R_c\sigma_f,
\label{rd2}
\end{equation}
over the range of variation of core radius $R_c$. Here $D_0$ is the
mean depth of field.

Clearly, these parameters also depend on the sample being studied.  In
our analysis, Eqs. (\ref{rd1}) and (\ref{rd2}) were fitted by a
maximum likelihood technique, and the resulting mean values of
$\sigma_f$ and $\sigma_s$ were accepted as the final estimates of the
structure parameters.  The difference of the estimates from
Eqs. (\ref{rd1}) and (\ref{rd2}) was included into the errors of the
final values.

\subsection{Methodological remarks}

For the idealized model considered above, the core-sampling method
would be expected to reproduce well the characteristics of the sample
under consideration. However, in reality, several factors may distort
the final results. For example, filaments may be apparently detected
as sheets, if they pass through the center of the core. Also the noise of
randomly distributed particles must be taken into account. Therefore,
a specific technique must be used to obtain stable results.

The analysis of the Las Campanas Redshift Survey (Doroshkevich et
al. 1996b) has shown that the stability of core-sampling with respect
to various accidental or systematic variations of density is very
high. During the sequential random rejection of particles, the final
estimates of mean separation of structure elements increased only as
$f^{-1/3}$. In general, the precision and stability of the
core-sampling method depends on the density contrast in structure
elements relative to the mean number density of points. In case of pure
Poissonian point distribution the 'structure' parameters have been
found with errors of about 50\% (Buryak, Doroshkevich \& Fong,
1994). Similar errors were found for the Soneira-Peebles model. On the
contrary, for the observed Las Campanas Redshift Survey errors were
found to be about 10\% only.

\begin{enumerate}
\item Sample preparation.

During the first step of analysis, when the sample is prepared, one
has to check whether the resulting 1D cluster distribution is
Poissonian as assumed. One has to ensure this requirement by
the variation of the radius of cylinder and the range of fit in
Eq. (\ref{lnN}).  During the further analysis the range of core radii
used for the fit of Eqs.  (\ref{rd1}) and (\ref{rd2}) is important to
get the two values in agreement.  These questions have to be solved
before the final analysis. This means that during each of the
intermediate steps leading to Figs. \ref{wallslice} to \ref{wallsND}, the
validity of all assumptions must be checked. Otherwise, the final
result will be meaningless.

\item The surface density of filaments.

The richness of a filament in the core depends both on the actual
properties of the point distribution analyzed and the geometry,
i.e. the position and orientation of the filament relative to the
core. Only the first piece of information is of interest; the geometry
can be characterized analytically and used to improve the stability of
the method.

The surface density of filaments $\sigma_f(f,\mu_{thr})$ depends on
the fraction of particles retained in the sample and the multiplicity
threshold.  The main characteristic of filamentary structure is the full
surface density of filaments, $\sigma_f(1,1)$. However, at this point
the noise is usually high and it is desirable to improve the estimate
of this value.

Basing on the geometrical model described above one can calculate the
distribution function for the length of intersection of identical
filaments with the core. Thus, calculating the dependence of both the
fraction $f$ and the surface density $\sigma_f$ on this length, we
find in linear approximation
\begin{equation}
\sigma_f(f)\approx \sigma_f(1)(1 - \sqrt{\frac{2}{3}}(1-f)^{1/2} + ...).
\label{sgm}
\end{equation}

This relation estimates the fraction of poor filaments for the ideal
case without noise. The linear fit of Eq. (\ref{sgm}) over some range
of $f$ allows one also to improve the estimate of $\sigma_f(1)$ for a
moderate noisy sample. However, in the case of very noise samples with
a significant fraction of filaments consisting only of one point, it
can be only used for small linking lengths, $l_c$, of the culling
procedure. Indeed, if $l_c\leq R_0$ the main part of noise particles
will be rejected (see Fig. 8, right).  Evidently, for smaller values
of $f$, the function $\sigma_f(f)$ resembles the actual properties of
the filaments (see, e.g. Doroshkevich et al.  1996b).

\item The linear density of sheets.

The richness of the elements of the sheet-like population both in
observational data and simulations is usually quite high. Thus, the
linear density of sheets can be estimated for a wide range of
parameters $f$ and $\mu_{thr}$, providing more reliable
results. However, even in this case the final results are sometimes
distorted because during reduction the fields are shortened as a result
of the appearance of empty edges. To avoid these distortions, a
correction procedure (see Doroshkevich et al. (1996b) was used.

The next problem is a filament population masquerading as sheets.
These apparent sheets are detected by the program if filaments cross
the cylinder near its axes. Obviously, this effect will be proportional
to the thickness of the filaments. The surface density $\sigma_s^a$ of
these apparent sheets is $\sigma_s^a = \pi \sigma_f r_f$, where $r_f$
is the radius of the filament (Buryak, Doroshkevich \& Fong,
1994). Usually the influence of this effect is negligible, but it can
become important for thick filaments. Moreover, apparent sheets can be
generated by the branch points (knots) of filamentary structure or
other clumps in the point distribution if their size is comparable to
the diameter of the core. The contribution of these knots can be
estimated as $\sigma_s^a = \pi n_{clm} r_{clm}^2$, where $ n_{clm}$
and $r_{clm}$ are respectively the volume density and the effective
radius of the clump. For a random network structure, the volume density
of the knots is approximately $n_{clm}\approx \sigma_f^{3/2}$.

\end{enumerate}

\section{Analysis of Voronoi-Tessellations}

\subsection{Construction of Voronoi tessellations}

We have used the code developed by R. Van de Weygaert (1991, 1994, see
also Goldwirth, da Costa, \& Van de Weygaert (1995)) in order to
construct Voronoi tessellations. Although the code has been developed
to allow for much more sophisticated cases, we limit ourselves to the
simplest case where a given number of points is distributed in walls
or filaments of a given thickness. To mimic noise we have added an
Poisson distribution of particles in the whole box. 

Using the same nuclei of the Voronoi tessellation, the filaments
(defined as the edges of the walls) would have approximately the same
separation as the walls (contrary to what we expect in nature).
Therefore, we did not combine a filamentary structure with a wall-like
structure in one realization, but made a second realization for
filaments alone. Note also that the filaments of a Voronoi
tessellation are not distributed like filaments of real
galaxies. Nevertheless the points of these Voronoi tessellation form
one-dimensional structure elements with a priori known properties
which should be recovered by the core sampling method.

The mean diameter of the Voronoi cells and the size of the box are
free input parameters of the code. For the core-sampling analysis it
is only important that the mean number of structure elements along a
random line crossing the box be large enough. We have chosen 6 cells
and 20 cells for the analysis of walls and filaments,
respectively. The analysis itself can be performed in units of the box
length. For the graphical representation we have chosen formally a box
size of 300 Mpc. We have chosen 200000 and 400000 points to represent
the walls and the filaments, respectively. Due to the formal
definition of the box size these numbers are not directly comparable
with the number of galaxies in the corresponding volume. However, in
that case the mean number density of points is of the order $10^{-2}$
Mpc$^{-3}$, i.e of the same order as the galaxy density. Also the the
mean separation of structure elements is of the same order as found
for galaxies, so that we can indeed draw conclusions from our test to
the situation found in observational surveys. The noise level chosen
(50\% for walls and 25\% for filaments) is in these realisations an
approximate upper limit for which the known properties can be
recovered with high accuracy.

\subsection{Wall distribution}

For the first test, we chose a mean size of the cells of 50 Mpc.
Our first realization consists of $6^3$ randomly distributed cells in box
of 300 Mpc length.  200 000 points are distributed only in walls
around the cells. We chose a Gaussian density profile of the
walls with a thickness of 2 Mpc. On the left hand side of
Fig. \ref{wallslice}, a slice of 8 Mpc thickness of this realization is
shown. In order to test the stability of the core-sampling method, we
made a second realization with 100 000 additional points which were
randomly distributed (Fig.  \ref{wallslice}, right).

\begin{figure*}[ht]
\epsfysize=5 cm
\vspace{1cm}
\caption{Top: a slice of 8 Mpc thickness, bottom: radial particle
distribution in the four cylinders shown at the top part of the figure
(the dashed line indicates the minimum core radius used in the
analysis), left: only walls, right: walls with a background of
randomly distributed particles }
\label{wallslice}
\end{figure*}

As explained in Sect. 2.1, the starting point of the core-sampling
analysis is the construction of randomly distributed cylinders in the
simulation. For illustration, we put four of these cylinders into each
of the slices shown in Fig.  \ref{wallslice}.  The diameter of the
cylinders is equal to the thickness of the slice (8 Mpc). In the lower
part of the figures, the distribution of points in these cylinders is
plotted vs.  the radius of the cylinder. This radius will be used in
the following as one diagnostic parameter as described in Sect.
2. Indeed, for sheet-like structures like the walls in our Voronoi
tessellations, the mean number of clusters in the 1D cluster analysis
is independent of the core radius. One can clearly see from
Fig. \ref{wallslice} that in both cases almost all walls will be
identified as walls.

Next, all particles $N_a$ in a given cylinder of radius $R_c$ are
projected onto the axis of the cylinder. From Fig. \ref{wallslice} it
is clear that there is also noise besides real structures in the
resulting one-dimensional point distribution. The degree of noise
depends on the chosen parameters of the generated sheets and on the
parameters of the core.  The noise particles in underdense regions can
be removed using the reduction procedure described in Sect. 2.1. The
reduced sample at which the further analysis will be performed is
characterized by three parameters: the fraction $f$, the threshold
$\mu_{thr}$, and the core radius $R_c$.

\begin{figure*}[h]
\epsfxsize=18 cm
\epsfbox{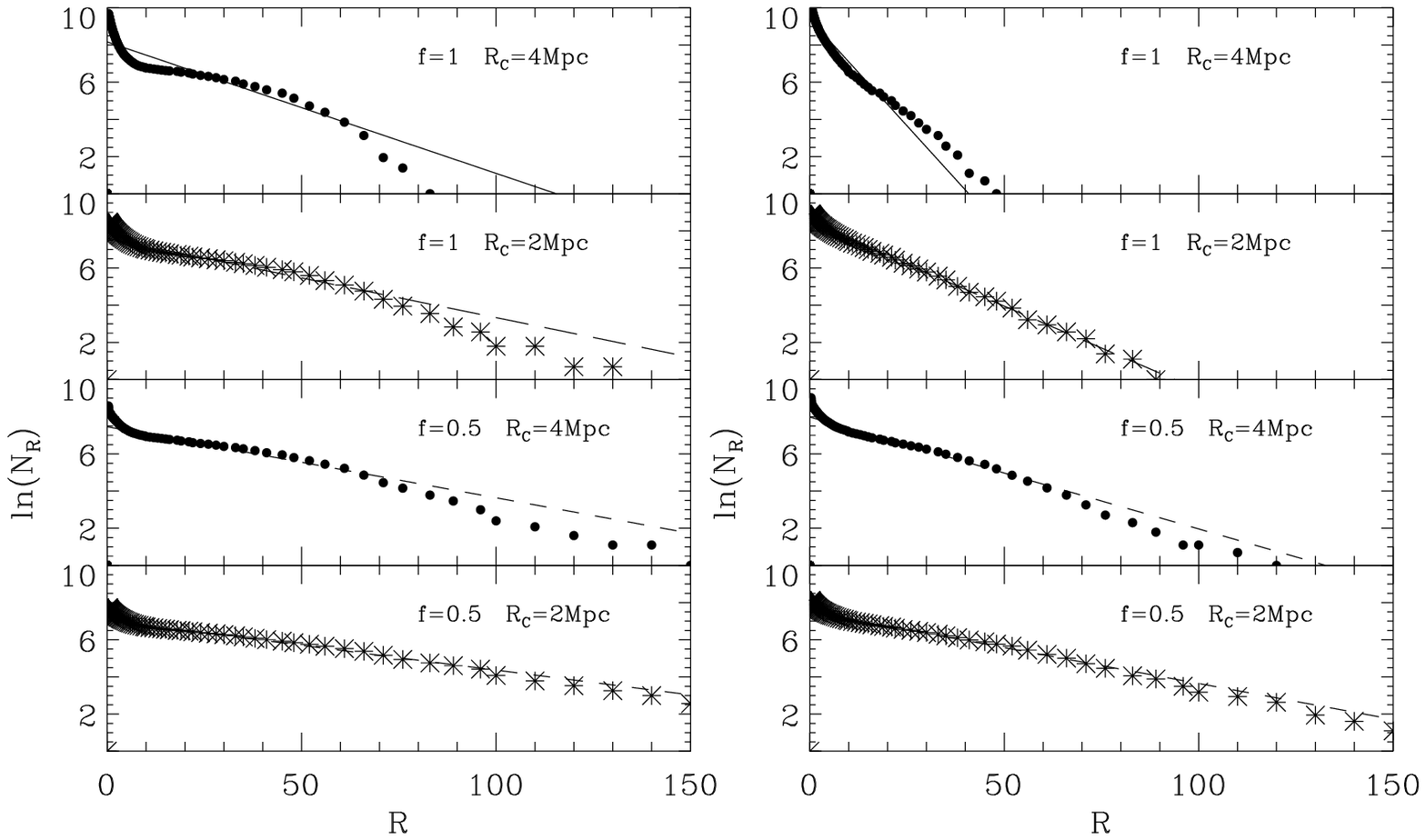}
\vspace{1cm}
\caption{A fit to the equation $\ln(N_R) = \ln(N_0) - R/R_0$ (left:
only walls, right: with a background)}
  \label{wallsNR}
\end{figure*}

Further, we use the iterative fitting procedure described in Sect. 2.2
to find the best fit to Eq. (\ref{lnN}). According to our assumption,
the resulting clusters must be Poisson distributed for the some range
of linking length.  However, this assumption must be tested for the
samples under consideration. As an example we show in
Fig. \ref{wallsNR} the fit to Eq.  (\ref{lnN}) for four sets of
parameters ($f=1$, 0.5, $r= 4 $ Mpc, 2 Mpc and $\mu_{thr}$ = 2: left
for walls and right for walls with randomly distributed points).

\begin{figure*}[h]
\epsfxsize=18 cm
\epsfbox{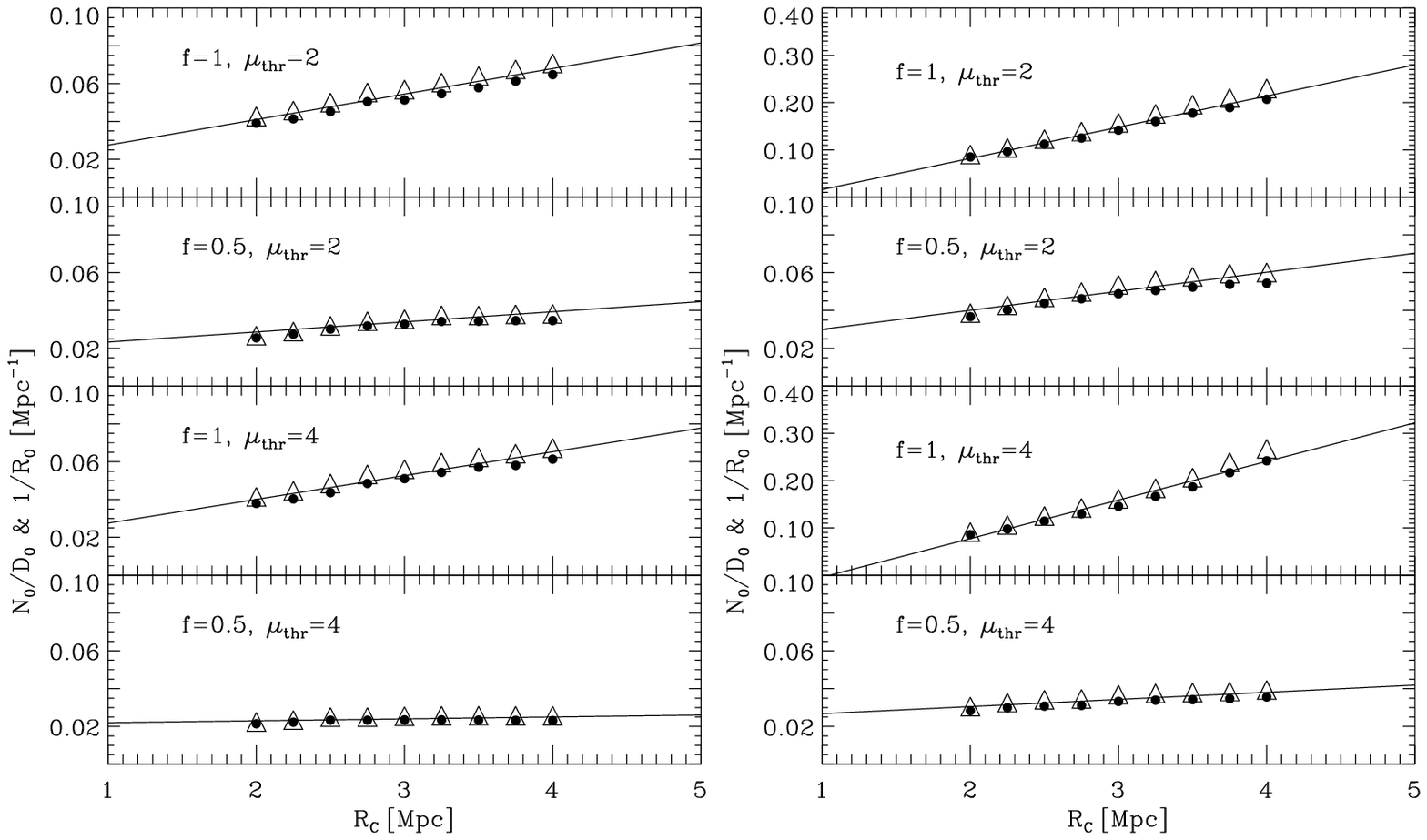}
\vspace{1cm}
\caption{A fit to the Eqs. (5) and (6) from which  $N_0/D_0)$ (dots) and
$1/R_0$ (triangles)
are obtained, (left: only walls, right: with a background)
}
\label{wallsND}
\end{figure*}

Using Eqs. (\ref{rd1}) and (\ref{rd2}), where $R_c$ is the diagnostic
parameter, we determine now the surface density of filaments and the
linear density of walls for each pair of parameters $f$ and
$\mu_{thr}$. This procedure is illustrated by Fig. \ref{wallsND}. for
four pairs of parameters.  (Note that each curve of
Fig. \ref{wallsNR} corresponds to one point in Fig. \ref{wallsND}.)
From Fig. \ref{wallsND} we extract the linear density of walls as the
zero point of the curve and the surface density of filaments as the
gradient. The fits to Eqs. (\ref{rd1}) and (\ref{rd2}) were performed
over 9 equally spaced values of $R_c$ in the interval 2 Mpc $ \leq
R_{c} \leq $ 4 Mpc.

\begin{figure*}[h]
\epsfxsize=18 cm
\epsfbox{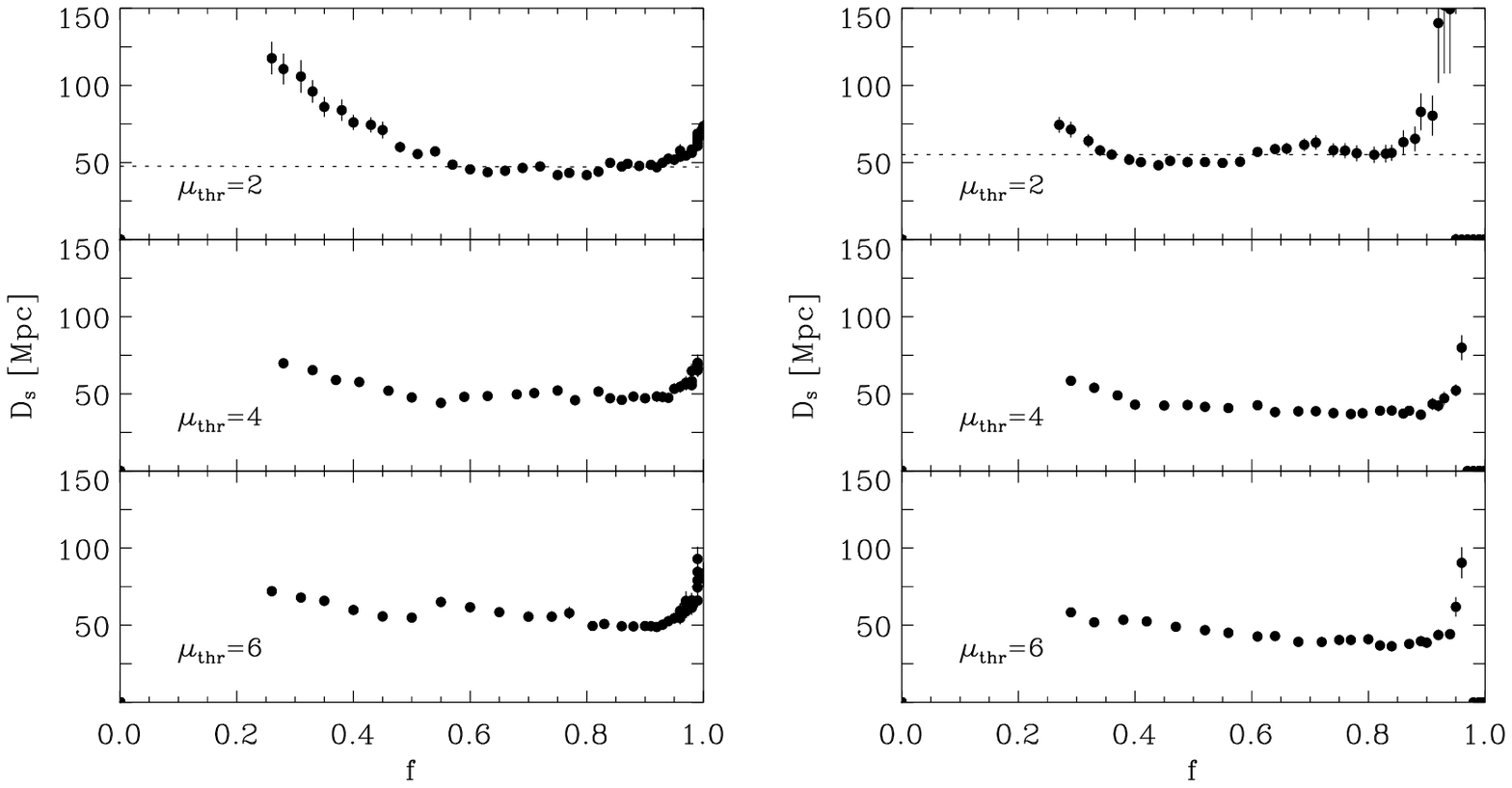}
\vspace{1cm}
\caption{Mean separation of walls for three multiplicity  thresholds (left:
only walls, right: with  background particles)
}
\label{wallswalls}
\end{figure*}

\begin{figure*}[h]
\epsfxsize=18 cm
\epsfbox{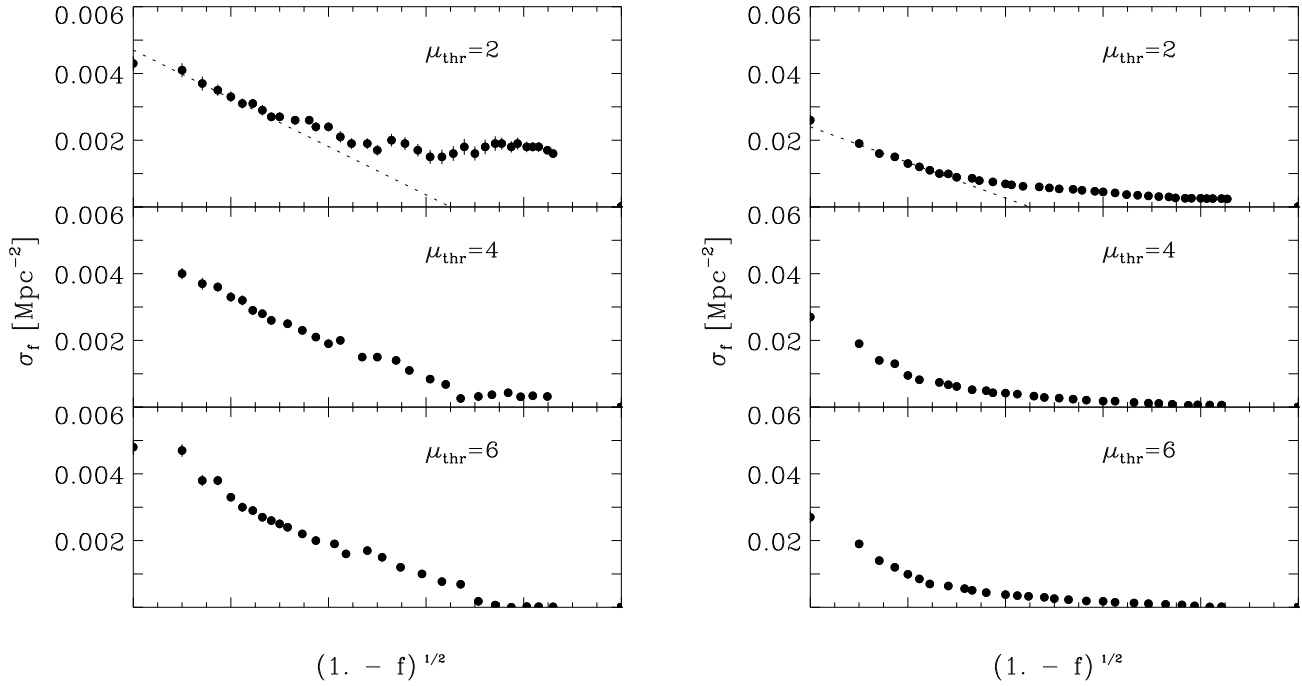}
\vspace{1cm}
\caption{Mean surface density of apparent filaments for three multiplicity
thresholds (left:only walls, right: with background particles)
}
\label{wallsfil}
\end{figure*}

As the final result of the analysis, we present in
Fig. \ref{wallswalls} the mean distance of the walls $D_s =
\sigma_s^{-1}$ and in Fig.  \ref{wallsfil} the surface density of
filaments $\sigma_f$ depending on the threshold $\mu_{thr}$. (Note,
that each curve of Fig. \ref{wallsND} provides one point in Fig.
\ref{wallswalls} and one point in Fig. \ref{wallsfil}.) The final
values of $\sigma_s$ and $\sigma_f$ are listed in Table \ref{walls}.

\begin{table*}[h]
\caption{
Test of the wall distribution.}
\label{walls}
\begin{tabular}{ccccc}
\hline
 $\mu_{thr}$ & $\sigma_s$       & $\sigma_f $      &$ \sigma_s^{noise}$&$
\sigma_f^{noise}$\\
             &$10^{-2} \Mpc^{-1}$&$10^{-2} \Mpc^{-2}$&$10^{-2} \Mpc^{-1}$ &
$10^{-2} \Mpc^{-2}$\\
\hline
    2        & $2.1\pm 0.29$    &  $0.43\pm0.21$    & $1.8\pm 0.20$     &
 $2.7\pm0.16$     \\
    4        & $2.0\pm 0.13$    &  $0.43\pm0.19$    & $2.5\pm 0.29$     &
 $2.7\pm0.16$     \\
    6        & $1.8\pm 0.16$    &  $0.48\pm0.21$    & $2.1\pm 0.17$     &
 $2.7\pm0.16$     \\
\hline
\end{tabular}
\end{table*}

The noise of apparent filaments is an objective characteristic of the
sample under consideration. The population of filaments dominates for
all thresholds $\mu_{thr}$ that resemble the small density contrast in
the walls, since the number of particles concentrated into sheets
exceeds the noise particles only by a factor of two. In this respect
our synthetic models differ from the Las Campanas Redshift Survey,
where for high threshold multiplicity the filament population had
disappeared, and only the richest sheet-like elements survived.

\subsection{Filament distribution}

As a second test we have investigated a pure filamentary structure and
filaments superimposed by randomly distributed particles. To this end
we have constructed a Voronoi tessellation with 400 000 particles
which are distributed at the edges of the cells, and a second one with
an additional 100 000 randomly distributed particles.  For this test
we chose a mean size of the cells of 15 Mpc. In Fig. \ref{filslice} we
show a slice of 4 Mpc thickness and four cores of radius 2 Mpc.
Further steps in the analysis are as above. We made two realisations
of Voronoi tessellations containing only filamentary structure: In the
first realisation we chose a radius of filaments of 0.2 Mpc, in the
second 0.1 Mpc.

For these synthetic models, we omit the figures corresponding to
Figs. \ref{wallsNR}-\ref{wallsND}, but present only the final results,
i.e.  the density of walls $\sigma_s(f)$ and filaments $\sigma_f(f)$
vs.  $(1-f)^{1/2}$ for two thresholds $\mu_{thr} = 2,~4$ and the two
different radii of filaments. (see
Figs. \ref{filswalls}-\ref{filsfils}). The final structure parameters
are given in Table \ref{filaments}.

\begin{figure*}[ht]
\epsfysize=5 cm
\vspace{1cm}
\caption{Top: a slice of 4 Mpc thickness, bottom: radial particle
distribution in the four cylinders shown at the top part of the figure
(the dashed line indicates the minimum core radius used in the
analysis), left: only filaments, right: filaments with a background of
randomly distributed particles }
\label{filslice}
\end{figure*}

\begin{figure*}[h]
\epsfxsize=18 cm
\epsfbox{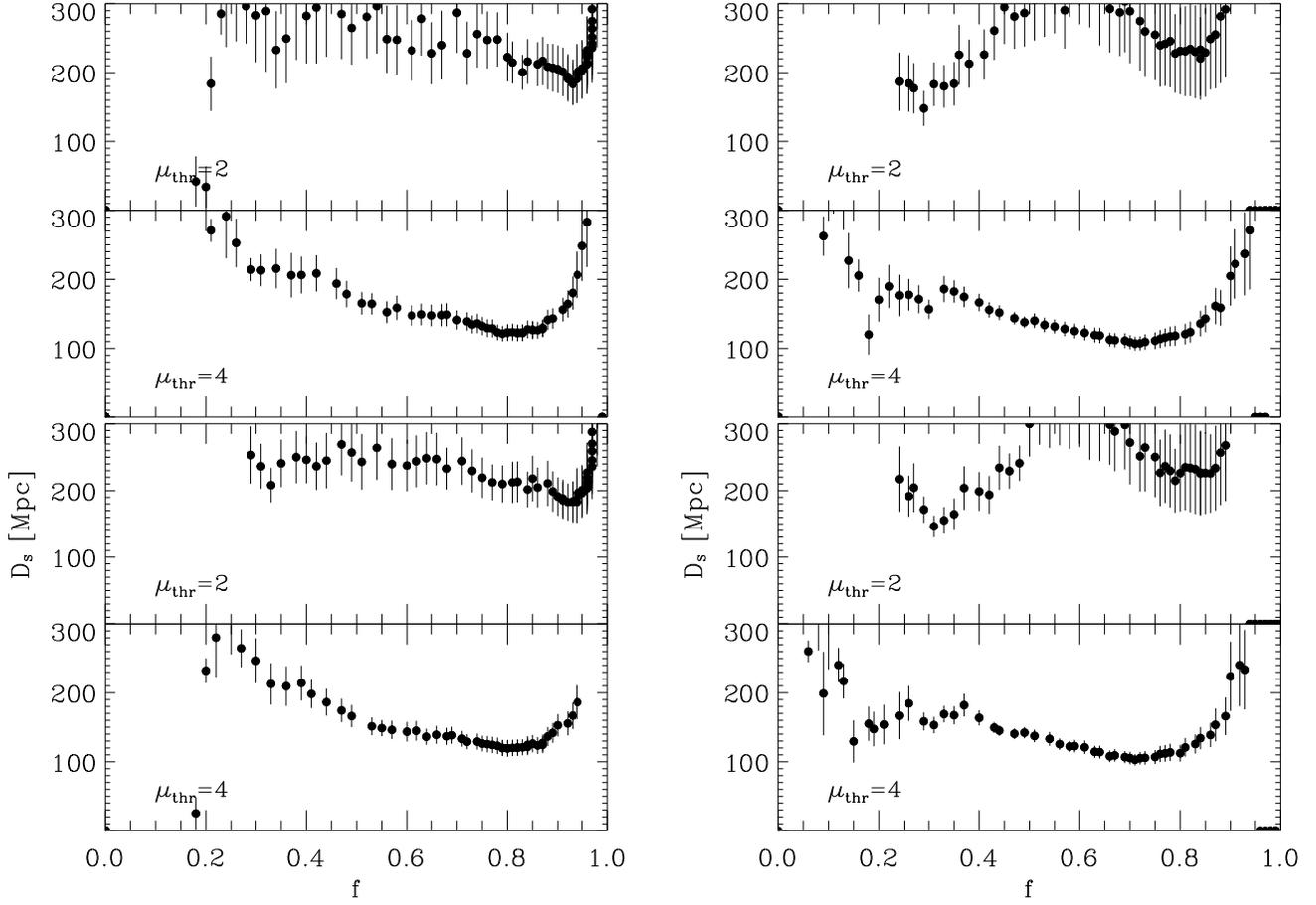}
\vspace{1cm}
\caption{Mean separation of apparent walls for two multiplicity
thresholds (top: radius of filaments 0.2 Mpc, bottom: radius of
filaments 0.1 Mpc, left: only filaments, right: with background
particles)
}
\label{filswalls}
\end{figure*}

\begin{figure*}[h]
\epsfxsize=18 cm
\epsfbox{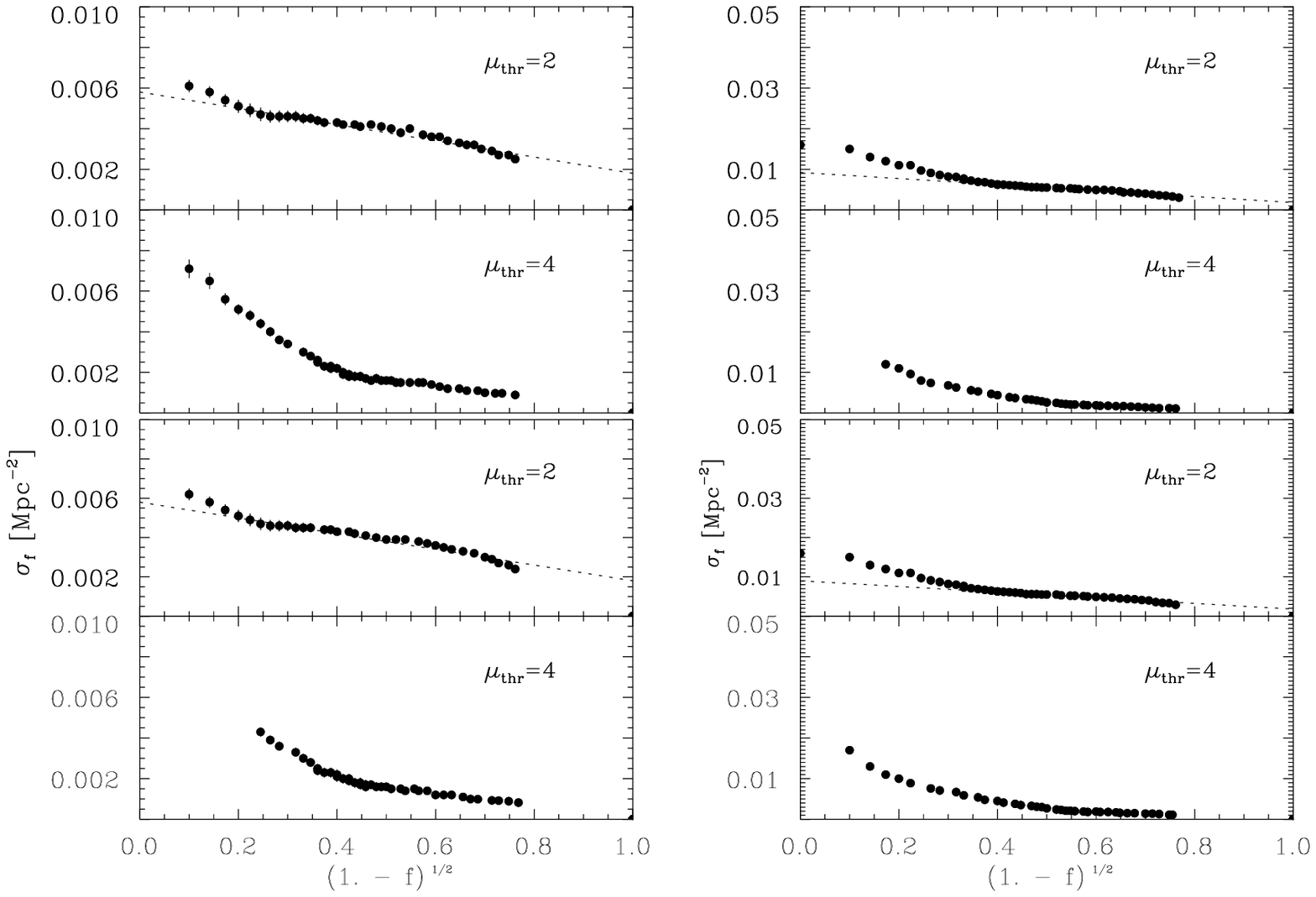}
\vspace{1cm}
\caption{Mean surface density of filaments for two multiplicity
thresholds (top: radius of filaments 0.2 Mpc, bottom: radius of
filaments 0.1 Mpc, left: only filaments, right: with background
particles) }
\label{filsfils}
\end{figure*}

Let us note that in this case the mean edge separation along a random
straight line is not identical to the mean size of the cell but is of
the order of the mean length of the filaments. The mean distance of
particles along the filaments in our sample is of the order of 1 Mpc.
Therefore, we expect that the set of particles forms a broken network
structure which can be characterized by the surface density of
filaments $\sigma_f(f)$. With a higher mean density we could create
also a complete network structure which, however, in reality does not
exist.

Since all filaments were created using the same procedure their
parameters can vary only statistically. We can use Eq. (\ref{sgm}) to
determine $\sigma_f(1)$ as is shown in Fig. \ref{filsfils} (dashed
line).  In case of $\mu_{thr} =4$ the points corresponding to a large
fraction $f$ do not follow the line predicted by Eq. (\ref{sgm}).
Rejecting these points, a linear fit would predict a somewhat smaller
$\sigma_f$. We expect this result, because we are now measuring only
the rare high-density filaments.

The noise strongly increases the full surface density of filaments
$\sigma_f(1)$ (Fig. \ref{filsfils}, right). The rejection of (10 -
15)\% of particles depresses the noise impact and leads us back to
noiseless data. Since most or all observed galaxies are incorporated
into structure elements the noise impact is not expected to be severe
in catalogues.  Following the rejection procedure we can obtain for
all realizations the same result $\sigma_f(1) = 0.0054$ which
corresponds to 13.6 Mpc mean separation (see Fig. \ref{filsfils}).

Fig. \ref{filslice} (bottom) indicates that some of the filaments
masquerade as sheets. We have discussed this problem already in
Sect. 2.4.  In this case the error bars strongly increase as a result
of the sparse sample statistics of elements which the core-sampling
method wrongly detects as sheets (see Fig. \ref{filswalls}).

\begin{table*}[h]
\caption{
Test of the filament distribution.}
\label{filaments}
\begin{tabular}{ccccc}
\hline
 $\mu_{thr}$ & $\sigma_s$       & $\sigma_f $      &$ \sigma_s^{noise}$&$
\sigma_f^{noise}$\\
             &$10^{-3} \Mpc^{-1}$&$10^{-2} \Mpc^{-2}$&$10^{-3} \Mpc^{-1}$ &
$10^{-2} \Mpc^{-2}$\\
\hline
 & & r=0.2 Mpc& &\\
\hline
    2        & $4.4\pm 0.8$    &  $0.57\pm0.02$    & $5.3\pm 1.0$    &
 $0.92\pm0.05$     \\
    4        & $7.3\pm 0.8$    &  $0.30\pm0.02$    & $6.5\pm 0.6$    &
 $0.50\pm0.04$     \\
\hline
 & & r=0.1 Mpc& &\\
\hline
    2        & $4.5\pm 0.9$    &  $0.58\pm0.01$    & $3.5\pm 1.2$    &
 $0.90\pm0.05$     \\
    4        & $6.1\pm 0.8$    &  $0.31\pm0.01$    & $5.5\pm 1.0$    &
 $0.49\pm0.04$     \\
\hline
\end{tabular}
\end{table*}

\section{Conclusions}

We have constructed a series of Voronoi tessellations with different
input parameters. Our realisations consist either only of structure
elements or of structure elements and noise. We have chosen a number
of points comparable with the situation in observational samples and
simulations of large scale structure formation, i.e. our realisations
correspond to models with all galaxies within structure elements and
models with additional randomly distributed galaxies.

In the first two tessellations, only walls were constructed. Using the
core-sampling method, we detected walls with a mean separation of 50
Mpc (see Table \ref{walls} and Fig. \ref{wallsNR}) in these synthetic
models. This corresponds exactly to the input parameter (the mean void
size) of the Voronoi tesselation. The method provided an accurate
reconstruction despite the additional noise which was added to the
tessellation (Fig.\ref{wallswalls} , right).  On the other hand, the
scales of the spurious filaments in the tessellation
(Fig. \ref{wallsfil}) are sensitive to the addition of noise. The
specific dependence found appears to be a property of the particular
synthetic model generated.

 The next four tessellations are all characterized by a mean cell size
of 15 Mpc. The cells are surrounded only by filaments.  Two
tessellations were made with filaments of radius 0.2 Mpc, two more
with 0.1 Mpc.  Independent of the radii of filaments, we were able to
determine the same mean distance of filaments (13.6 Mpc); however,
this estimate was slightly reduced by adding randomly distributed
points as background (cp. Fig. \ref{filsfils} , dashed lines). The
detected spurious walls have distances of the order of the box size
and very large errors so that they can be easily ruled out.

Our results demonstrate clearly that the core-sampling method allows
one not only to determine the structure elements, but also to measure
their characteristic mean separation or density almost independently
of the influence of a substantial noise component.  From the good
agreement between the expected and recovered structure parameters, we
conclude that the core-sampling method is a powerful tool for further
investigations of observational surveys and the determination of
structure elements such as filaments and walls of galaxies. Recently,
the core-sampling method has been successfully applied both to
numerical simulations (Doroshkevich et al. 1996a) and the Las Campanas
galaxy sample (Doroshkevich et al. 1996b).

{\bf Acknowledgements:} We would like to thank Rien Van de Weygaert
for allowing us to use his code for generating Voronoi tessellations
and Ron Kates for valuable discussion of the manuscript. This paper
was supported in part by Danmark Grundforskningsfond through its
support for the establishment of the Theoretical Astrophysics
Center. A.G.D. was partly supported by the INTAS grant 93-0068 and by
the Center for Cosmoparticle Physics 'Cosmion' in the framework of the
project 'Cosmoparticle Physics'. S.G. wishes to express gratitude for
the hospitality of the TAC Copenhagen.


\end{document}